\documentclass{emulateapj}
\usepackage{amsmath}
\usepackage[]{natbib}
\usepackage{graphics}
\usepackage{threeparttable}
\usepackage{apjfonts}

\newcommand{\etal}{et al.}  
\newcommand{\per}{\ensuremath{^{-1}}}

\newcommand{\hal}{H\ensuremath{\alpha}}
\newcommand{\hbeta}{H\ensuremath{\beta}} 
\newcommand{\hgamma}{H\ensuremath{\gamma}} 
\newcommand{\hdelta}{H\ensuremath{\delta}} 

\newcommand{\hst}{\emph{HST}}

\newcommand{\lsun}{\ensuremath{{L}_{\odot}}} 
\newcommand{\kms}{km~s\ensuremath{^{-1}}}

\newcommand{\sigmasixtyeight}{\ensuremath{\sigma_{68}}}

\usepackage{color}

\begin{document}

\shorttitle{Type 2 Quasar Variability}
\shortauthors{Barth \etal}

\title{A Search for Optical Variability of Type 2 Quasars in SDSS Stripe 82}

\author{Aaron J. Barth\altaffilmark{1},
  Alexey~Voevodkin\altaffilmark{2,3}, Daniel
  J. Carson\altaffilmark{1}, Przemys\l aw Wo\'zniak\altaffilmark{3}}

\altaffiltext{1}{Department of Physics and Astronomy, 4129 Frederick
  Reines Hall, University of California, Irvine, CA, 92697-4575, USA;
  barth@uci.edu}

\altaffiltext{2}{Space Research Institute (IKI), Profsoyuznaya 84/32,
  Moscow, Russia, 117997 }

\altaffiltext{3}{Los Alamos National Laboratory, MS-D466, Los Alamos, NM, 87545}

\begin{abstract}

Hundreds of Type 2 quasars have been identified in Sloan Digital Sky
Survey (SDSS) data, and there is substantial evidence that they are
generally galaxies with highly obscured central engines, in accord
with unified models for active galactic nuclei (AGNs). A
straightforward expectation of unified models is that highly obscured
Type 2 AGNs should show little or no optical variability on timescales
of days to years. As a test of this prediction, we have carried out a
search for variability in Type 2 quasars in SDSS Stripe 82 using
difference-imaging photometry.  Starting with the Type 2 AGN catalogs
of Zakamska \etal\ (2003) and Reyes \etal\ (2008), we find evidence of
significant $g$-band variability in 17 out of 173 objects for which
light curves could be measured from the Stripe 82 data. To determine
the nature of this variability, we obtained new Keck
spectropolarimetry observations for seven of these variable AGNs.  The
Keck data show that these objects have low continuum polarizations
($p\lesssim1\%$ in most cases) and all seven have broad \hal\ and/or
\ion{Mg}{2} emission lines in their total (unpolarized) spectra,
indicating that they should actually be classified as Type 1 AGNs.  We
conclude that the primary reason variability is found in the
SDSS-selected Type 2 AGN samples is that these samples contain a small
fraction of Type 1 AGNs as contaminants, and it is not necessary to
invoke more exotic possible explanations such as a population of
``naked'' or unobscured Type 2 quasars.  Aside from misclassified Type
1 objects, the Type 2 quasars do not generally show detectable optical
variability over the duration of the Stripe 82 survey.

\end{abstract}



\section{Introduction}
\label{sec:intro}

Unified models of active galactic nuclei (AGNs) have been extremely
successful at explaining the different appearance of Type 1
(broad-lined) AGNs and Type 2 AGNs (those lacking broad emission
lines) as resulting from anisotropic obscuration in a dusty torus
surrounding the AGN's central engine and broad-line region (BLR).  The
simplest version of the unified model \citep{antonucci1993} posits
that all Type 2 AGNs contain a hidden Type 1 nucleus that is obscured
along our line of sight by a toroidal dust distribution.  A broad
range of observational evidence supports the general picture of
toroidal obscuration regions in AGNs, including detection of
``hidden'' broad emission lines seen in polarized scattered light
\citep{antonucci1985}, high X-ray obscuring columns in many Type 2
Seyferts \citep{mulchaey1992}, and ionization cones in the narrow-line
regions of Seyfert 2 galaxies \citep{pogge1988}.  Recent work has
suggested that orientation alone is probably insufficient to explain
all of the differences between Type 1 and Type 2 AGNs, and in
particular, the distribution of dust torus covering factors may be
different for Type 1 and Type 2 samples \citep{ramos-almeida2011,
  elitzur2012}.  Nevertheless, even if the Type 1/2 dichotomy is not
purely the result of orientation differences, it is firmly established
that many Type 2 AGNs do contain a hidden Type 1 nucleus. For samples
of Seyfert 2 galaxies at low redshift, the fraction found to contain
hidden broad-line regions in polarized light is between $\sim35\%$ and
50\% \citep{moran2000, tran2001}.

The Sloan Digital Sky Survey \citep[SDSS;][]{york2000} provided the
first opportunity for identification of large numbers of
optically-selected Type 2 quasars.  \citet[][hereinafter
  Z03]{zakamska2003} carried out the first systematic search using
SDSS Data Release 1 \citep[DR1;][]{abazajian2003}, and identified 291
Type 2 AGNs of which about half had [\ion{O}{3}] luminosities above
$3\times10^8$ \lsun, placing them in the same luminosity range as
quasars. Extending this work further using SDSS DR6,
\citet[][hereinafter R08]{reyes2008} used slightly modified selection
criteria to compile a sample of 887 Type 2 AGNs all with [\ion{O}{3}]
luminosities above $2\times10^8$ \lsun, and identified these as Type 2
counterparts of luminous quasars. Follow-up X-ray observations
demonstrated that many of these objects are moderately to highly
absorbed \citep{zakamska2004, vignali2004, vignali2006, jia2012}, and
\citet{zakamska2005} found evidence of high optical continuum
polarization and polarized broad emission lines in several
objects. Such properties support the identification of these objects
as obscured Type 2 versions of quasars.

Temporal variability is one of the hallmarks of AGN activity, but for
highly obscured AGNs the unified model would predict that optical
variability should be highly suppressed since the observer's line of
sight to the central continuum source and broad-line region is blocked
by the dusty torus.  Even if some AGN continuum emission was visible
in scattered light, the dominant and non-variable contribution of host
galaxy starlight would strongly dilute the variability of the
scattered AGN light.  Additionally, any variability in the scattered
light component would be temporally blurred by reflection from a
spatially extended scattering region, so that short-term flux
variations on timescales shorter than the light-crossing time across
the scattering region would tend to be smoothed out.  Given these
considerations, optical variability of Type 2 AGNs has seldom been
investigated, although some variability surveys have identified a
small number of candidate variable Type 2 AGNs
\citep[e.g.,][]{bershady1988, sarajedini2006}.  \citet{yip2009}
carried out the largest systematic investigation of Type 2 AGN
variability to date.  They used SDSS spectroscopic data to search for
variability in the subset of AGNs having two epochs of spectroscopy,
and found no evidence of continuum or emission-line variability in
Type 2 AGNs on timescales of months to a few years.

One intriguing result comes from a long-term monitoring program by
\citet{hawkins2004}.  In a 25-year photographic survey,
\citet{hawkins2004} found evidence of optical flux variations in some
AGNs which had spectra consistent with a Type 2 classification.
Overall, 10\% of the emission-line galaxies in the \citet{hawkins2004}
survey were found to be Type 2 AGNs showing significant flux
variability.  Hawkins' interpretation was that these objects do not
conform to the standard unified model. He proposed that these are
narrow-lined AGNs that are \emph{not} obscured and which do not
possess broad-line regions at all, i.e., ``naked AGNs'' in which the
central engine is unobscured.  While plausible candidates for naked
Type 2 nuclei (sometimes elsewhere referred to as ``true'' Type 2
AGNs) have been identified at very low luminosities in nearby galaxies
\citep[e.g.,][]{pappa2001, shi2010, tran2011}, it would be surprising
if such objects, totally lacking a BLR, were common at high
luminosity.  (See Antonucci 2012 for a critique of ``true'' Type 2 AGN
classification.)

The SDSS Stripe 82 database has enabled a variety of new
investigations of AGN variability.  Stripe 82 data has been used
extensively to examine continuum variability in quasars
\citep{sesar2007, bhatti2010, ai2010, macleod2010, meusinger2011,
  gu2011, sakata2011, voevodkin2011, schmidt2012, macleod2012,
  zuo2012} and to develop new methods for quasar selection by
variability \citep{schmidt2010, macleod2011, butler2011, palanque2011,
  wu2011}, but Type 2 quasar variability in Stripe 82 has not
previously been examined. The combination of the Z03 and R08 samples
together with the Stripe 82 data archive provides an opportunity to
investigate the photometric variability of a substantial sample of
Type 2 quasars over a duration of nearly a decade.  While there are
now extensive samples of obscured high-luminosity AGNs selected from
infrared and X-ray surveys, the SDSS-selected Type 2 quasars are
particularly well suited to examination of optical variability because
of their relatively bright optical magnitudes and the large number
($>200$) that are located in the Stripe 82 survey area.

In this paper, we describe a new search for variability in luminous
Type 2 AGNs in Stripe 82 using image-subtraction photometry, and
follow-up observations to examine the nature of the variable sources.
The methods for identification of variable AGNs are described in
Section \ref{sec:data}, along with illustrations of the light curves
and spectra of the selected objects.  Section \ref{sec:spol} describes
new Keck Observatory spectropolarimetry observations of seven objects.
Section \ref{individualobjects} describes properties of the individual
objects, and in Section \ref{sec:discussion} we discuss the possible
causes of variability in these objects. We find that over the duration
of the Stripe 82 survey, the majority of the Type 2 quasars do not
show any evidence of optical variability, consistent with
straightforward expectations from the unified model.  However, about
10\% of previously identified Type 2 AGNs in Stripe 82 do exhibit
significant flux variations.  Examination of the SDSS spectra and our
new Keck observations suggests that most of the variable objects are
in fact Type 1 AGNs that were previously classified as Type 2 objects
either due to low S/N in the SDSS data, lack of sufficient
wavelength coverage to detect broad \hal\ or \ion{Mg}{2} in the SDSS
spectra, or other spectral peculiarities that rendered the
classification ambiguous. Some appear to be variable intermediate-type
(1.8/1.9) AGNs or narrow-line Seyfert 1 galaxies with properties
similar to previously known examples of Type 2 quasar impostors.  We
cannot rule out the possibility that some objects identified in our
search may be genuinely variable Type 2 AGNs, but such objects, if
they exist at all, must be very rare.


\section{Data and Measurements}
\label{sec:data}

\subsection{Sample Selection}

We base our study on the Type 2 AGN samples of Z03 and R08. The
samples were compiled using SDSS Data Release 1
\citep[DR1;][]{abazajian2003} and Data Release 6
\citep[DR6;][]{adelman-mccarthy2008} spectroscopic data, respectively.
We briefly summarize the selection criteria of the two samples, and
refer the reader to Z03 and R08 for full descriptions of the sample
selection methodology.  The Z03 sample includes objects in the
redshift range $0.3 < z < 0.83$ having narrow emission lines with
ratios consistent with a high-ionization AGN classification and no
detectable broad-line components.  Z03 identified 291 Type 2 AGNs,
with [\ion{O}{3}] $\lambda5007$ luminosities spanning a broad range
from $\sim10^7$ to $10^{10}$ \lsun.  About half of the sample has
$L$([\ion{O}{3}])$>3\times10^8$ \lsun, and these objects can be
identified as luminous Type 2 quasars.  In the R08 catalog, Type 2
quasars were selected based on slightly modified criteria: redshifts
$z<0.83$ were selected (with no minimum redshift), a luminosity
threshold of $L$([\ion{O}{3}])$>10^{8.3}$ \lsun\ was applied in order
to exclude AGNs of lower luminosity (i.e., Seyfert 2 galaxies), and a
modified method was used to exclude objects showing broad components
on the Balmer emission lines.  Furthermore, the [\ion{O}{3}] line was
required to have an equivalent width $>4$ \AA\ and a S/N sufficient
for a clear AGN classification.  For objects above the [\ion{O}{3}]
luminosity threshold, R08 recover the selection of $>90\%$ of DR1
objects that were previously identified by Z03.  Those objects in the
Z03 catalog which are not in the R08 catalog are either below the
minimum [\ion{O}{3}] luminosity to be included, or they have low-S/N
spectra or could not be classified unambiguously as AGN.

A substantial fraction of these Type 2 quasars are found in the Stripe
82 region, enabling us to examine their light curves over a
$\sim9$-year duration.  Stripe 82 covers a strip of sky along the
celestial equator in the region $-60\arcdeg \le \alpha \le60\arcdeg$
and $-1.25\arcdeg \le \delta \le 1.25\arcdeg$
\citep{adelman-mccarthy2008}.  The observations of the Stripe 82 field
started on 1998 September 19 and continued each fall during a 2--3
month period. The median number of observations per field is about 70,
and the last 5 years of the light curves are more densely sampled than
the initial years.  The Stripe 82 region contains 110 Z03 AGNs and
143 R08 AGNs, respectively, with 45 objects in common between the two
samples.  Thus, we select 208 unique AGNs in Stripe 82 as our starting
sample.  For reasons described above, the R08 sample is expected to
have a higher purity for selection of genuine Type 2 quasars than the
Z03 sample, but for completeness, we start with the set of all of the
Z03 and R08 objects located in Stripe 82 to examine their variability.

\begin{figure}
  \includegraphics[width=\columnwidth]{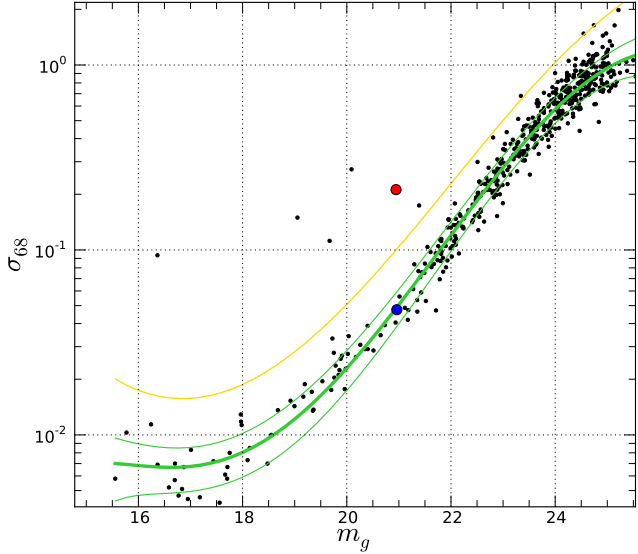}
  \caption{The quality of difference-image photometry for a typical
    field. Black points correspond to all sources detected in that
    field. The thick green line shows the median of the variability of
    the sources, and thin green lines show the $1\sigma$ scatter
    around the median.  The yellow line denotes the $5\sigma$ threshold
    used to select variable sources.  The large red data point
    represents one of the variable AGNs selected in this study, SDSS
    J033123, and the large blue point is a non-variable object having
    a similar median magnitude of $m_g \approx 21$ mag.  The light
    curves for these two sources are compared in Figure
    \ref{fig:examplelightcurve}.  
    \label{fig:example}}
\end{figure}

\subsection{Analysis of Variability}

Since the Type 2 AGNs are spatially extended sources and contain a
substantial or dominant contribution of starlight from the host
galaxies, the standard SDSS catalog photometry is not optimal for
detection of low-level variability.  Given the expectation that any
variable flux, if present at all, should be a weak and spatially
compact component superposed on the extended host galaxy, an
image-subtraction approach becomes very advantageous in this context.
Our examination of the AGN light curves is based on difference image
photometry \citep{alard1998} applied to the $g$-band Stripe 82 imaging
database. The details of the procedure are described by
\citet{voevodkin2011}. One of its outputs is a set of high quality
relative light curves for all sources in a given SDSS field.

\begin{figure}
  \includegraphics[width=\columnwidth]{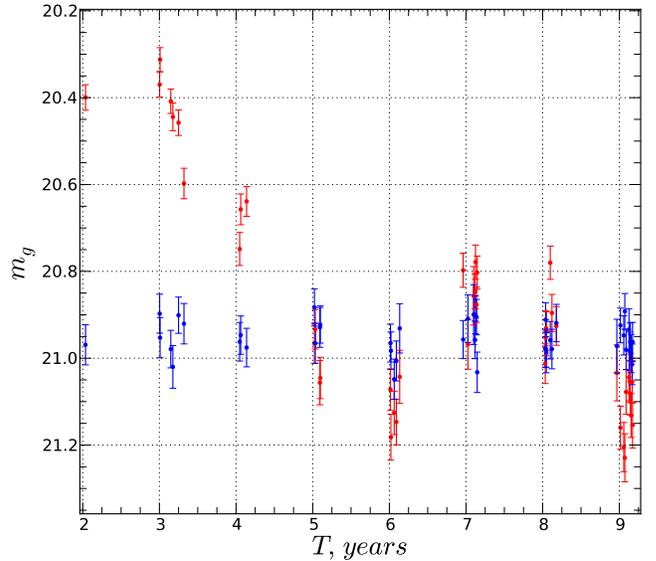}
  \caption{Light curves for the two objects represented by the red and
    blue points in Figure \ref{fig:example}, illustrating variable and
    nonvariable objects at similar median magnitudes of $m_g\approx21$
    mag.  The zeropoint of the time axis corresponds to 1998 September
    19.
    \label{fig:examplelightcurve}}
\end{figure}

The quality of photometry for a typical field is demonstrated in
Figure~\ref{fig:example}. It shows the root-mean-square (rms)
fluctuation level of the light curve (specifically, the 68th
percentile of absolute deviations from the median, \sigmasixtyeight)
versus median magnitude for every source identified in that field.
This includes both pointlike and extended objects, down to a detection
limit that corresponds to a $3\sigma$ threshold in a high-S/N
reference image constructed by co-adding individual images with the
best seeing. (The specific detection limit does not affect our results
because all of the Z03 and R08 AGNs we examine are at least a few
magnitudes brighter than this limit.) The vast majority of objects in
the field are non-variable, and fall in a tight diagonal locus in
which the increasing rms at faint magnitudes is the result of
photometric uncertainties.  Points that lie well above this locus are
objects showing significant variability in their light curves.
Figure~\ref{fig:examplelightcurve} illustrates sample light curves for
the two 21st-magnitude objects highlighted in Figure
\ref{fig:example}, one variable and one non-variable.

\begin{figure}
  \includegraphics[width=\columnwidth]{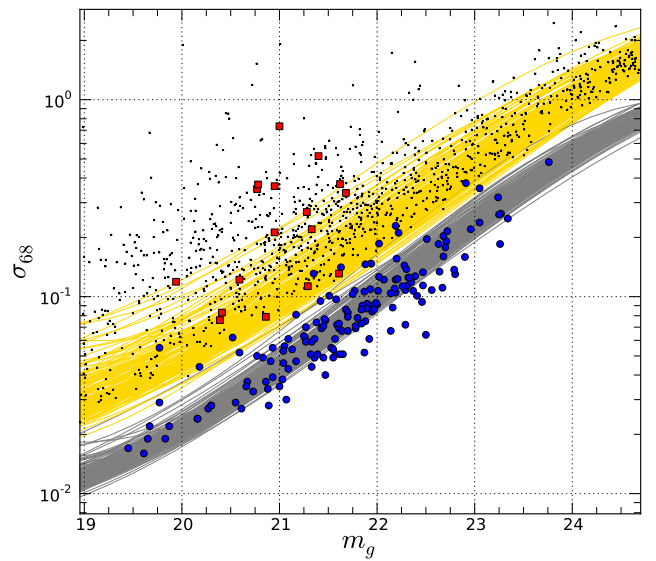}
  \caption{Variability selection for all Stripe 82 fields. Red squares and
    blue points correspond to Type 2 quasars lying above and below the
    variability threshold in their respective fields.  The sets of gray and
    yellow lines show the median loci and $5\sigma$ variability thresholds
    of all considered fields.  Small black points represent all sources
    which lie above the variability threshold for their field.
    \label{fig:var_src}
  }
\end{figure}

The data quality was sufficient to produce image-subtraction light
curves for 173 of the 208 AGNs.  The pipeline was unable to produce
light curves for the remaining 35 AGNs for a variety of reasons,
including locations close to the edges of images or close to very
bright stars, a lack of suitable calibration stars, or a location in
fields having fewer than 20 suitable epochs of good-quality imaging.
Of the 173 AGNs with light curves, 54 are from the Z03 sample only, 84
are from R08 only, and 35 are present in both samples.  These 173 AGNs
lie in 172 separate Stripe 82 fields, since one field contains two
AGNs.  The median number of points in the light curves is 36.

\begin{figure*}[t!]
\begin{center}
\scalebox{0.6}{\includegraphics{{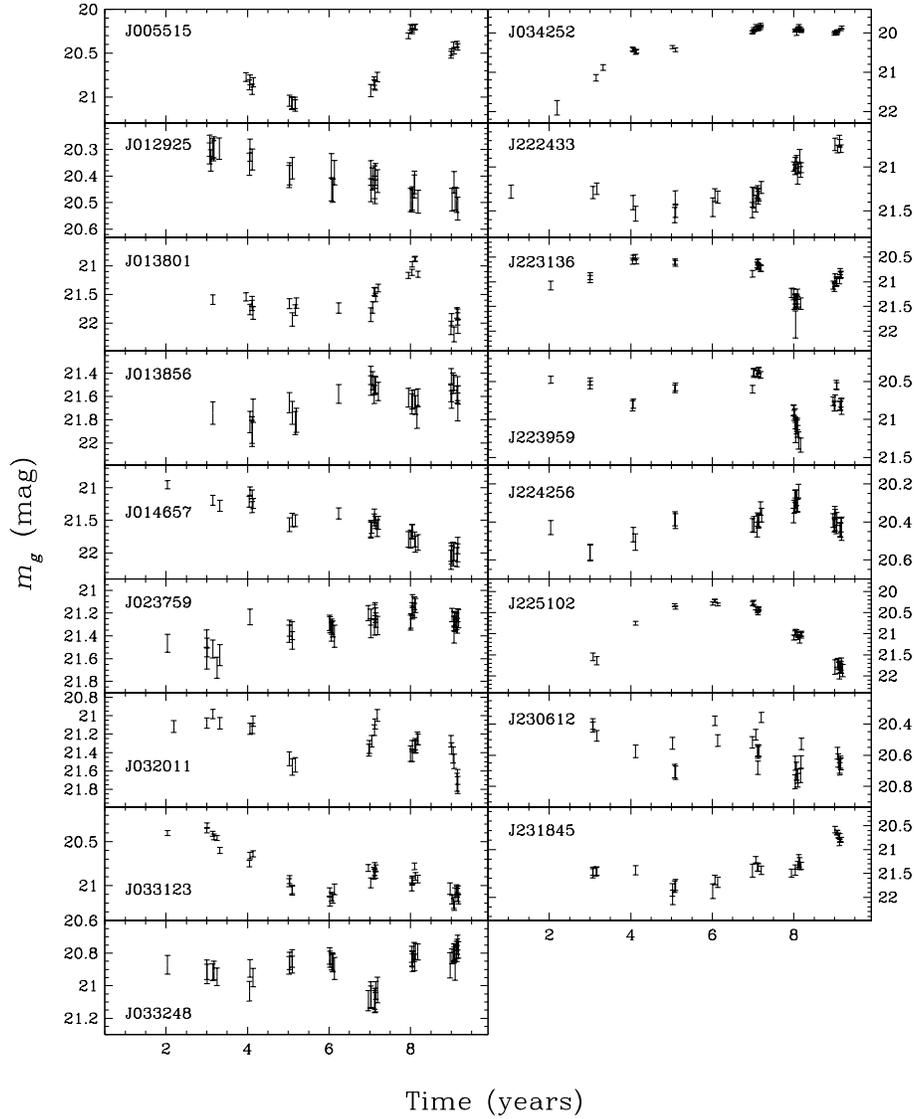}}}
\caption{Light curves for the 17 variable objects. 
\label{fig:lightcurves}}
\end{center}
\end{figure*}

\begin{deluxetable*}{lccccccc} 
  \tablecaption{Variable Type 2 AGN Candidates}
  \tablehead{
    \colhead{SDSS ID} & 
    \colhead{$\alpha$} &
    \colhead{$\delta$} &
    \colhead{Sample} &
    \colhead{$z$} & 
    \colhead{$m_g$ (mag)} & 
    \colhead{\sigmasixtyeight\ (mag)} & 
    \colhead{$\log L$([\ion{O}{3}])/\lsun}
  }
  \startdata
  J005515.81$-$004648.5 &     13.815917&  $-$0.780167 & Z &    0.345 & 20.77 &   0.352  & 8.15 \\
  J012925.81$-$005900.2 &     22.357571&  $-$0.983405 & R &    0.711 & 20.41 &   0.083  & 9.71 \\ 
  J013801.57$-$004946.5 &     24.506542&  $-$0.829583 & Z &    0.433 & 21.68 &   0.336  & 8.29 \\
  J013856.14$+$003437.4 &     24.733919&  $+$0.577056 & Z &    0.478 & 21.61 &   0.131  & 8.29 \\ 
  J014657.23$+$005537.2 &     26.738502&  $+$0.927000 & Z &    0.422 & 21.62 &   0.373  & 8.04 \\
  J023759.75$+$001723.5 &     39.499002&  $+$0.289889 & Z &    0.335 & 21.29 &   0.113  & 7.46 \\
  J032011.94$+$002702.2 &     50.049792&  $+$0.450611 & Z &    0.443 & 21.33 &   0.220  & 7.92 \\
  J033123.14$-$005930.7 &     52.846417&  $-$0.991889 & Z &    0.431 & 20.95 &   0.212  & 9.05 \\ 
  J033248.49$-$001012.3 &     53.202095&  $-$0.170094 & Z+R &  0.310 & 20.86 &   0.079  & 8.53 \\ 
  J034252.47$+$005252.4 &     55.718632&  $+$0.881232 & Z+R &  0.565 & 19.94 &   0.119  & 8.93 \\ 
  J222433.31$-$003634.0 &    336.138824&  $-$0.609475 & R &    0.588 & 21.28 &   0.269  & 8.87 \\
  J223136.27$-$011045.0 &    337.901126&  $-$1.179167 & Z &    0.436 & 20.95 &   0.364  & 8.60 \\
  J223959.04$+$005138.3 &    339.996002&  $+$0.860639 & Z &    0.384 & 20.78 &   0.371  & 8.15 \\
  J224256.47$+$005155.2 &    340.735382&  $+$0.865334 & R &    0.410 & 20.39 &   0.076  & 9.26 \\ 
  J225102.40$-$000459.8 &    342.760000&  $-$0.083306 & Z &    0.550 & 21.00 &   0.733  & 9.13 \\   
  J230612.90$-$005912.6 &    346.553772&  $-$0.986841 & R &    0.250 & 20.59 &   0.122  & 8.38 \\ 
  J231845.12$-$002951.4 &    349.688002&  $-$0.497611 & Z &    0.397 & 21.40 &   0.517  & 8.00 \\
  \enddata
  \tablecomments{We include both the sexagesimal and decimal
    coordinates (all J2000) to facilitate cross-matching with the original Z03 and
    R08 catalogs.  The ``Sample'' column indicates whether an object is in 
    the Z03 or R08 samples, or both.  The listed
    $g$-band magnitudes represent the median magnitude over the Stripe
    82 light curve.  
    [\ion{O}{3}] luminosities are taken from the Z03 and R08
    catalogs. The SDSS sexagesimal designations are those listed in
    NED for each object, and we note that for some objects the right 
    ascension and declination coordinates
    differ in the last decimal places from the object designations
    listed in Z03.}
  \label{tab:var_agns}
\end{deluxetable*}

In order to identify variable AGNs, we require criteria to select
objects having light curve variations significantly greater than the
expected scatter due to photometric errors for non-variable objects of
a given magnitude.  We use the \sigmasixtyeight\ vs.\ $m_g$ diagram
(as in Figure \ref{fig:example}) for each photometric field as the
basis of our selection.  The non-variable locus can be characterized
by some function of magnitude and by the scatter around this function,
where the scatter also depends on magnitude. To find an analytic
approximation for the non-variable locus, we first calculated the
median value of \sigmasixtyeight\ and its standard deviation over a
running window containing 10 data points ranked in order of increasing
$m_g$. Any $2\sigma$ or greater outliers were then excluded and the
median values were recalculated, producing a smoothed set of data
points which were fit in log-log space by a 4th-order polynomial.
This polynomial fit defines the non-variable locus illustrated with a
thick green curve in Figure \ref{fig:example}.  To determine the
$1\sigma$ scatter band, we first flattened the data by subtracting the
analytic approximation for the non-variable locus from the data
points.  Then, in the flattened data, we calculated the standard
deviation of \sigmasixtyeight\ over a running boxcar containing 10
data points, producing a smoothed version of the scatter region.  This
set was then fitted with a 3rd-order polynomial in log-log space, and
transformed back to the \sigmasixtyeight\ vs.\ $m_g$ plane as
illustrated in Figure \ref{fig:example} with thin green curves.

\begin{figure*}[t!]
\begin{center}
\scalebox{0.6}{\includegraphics{{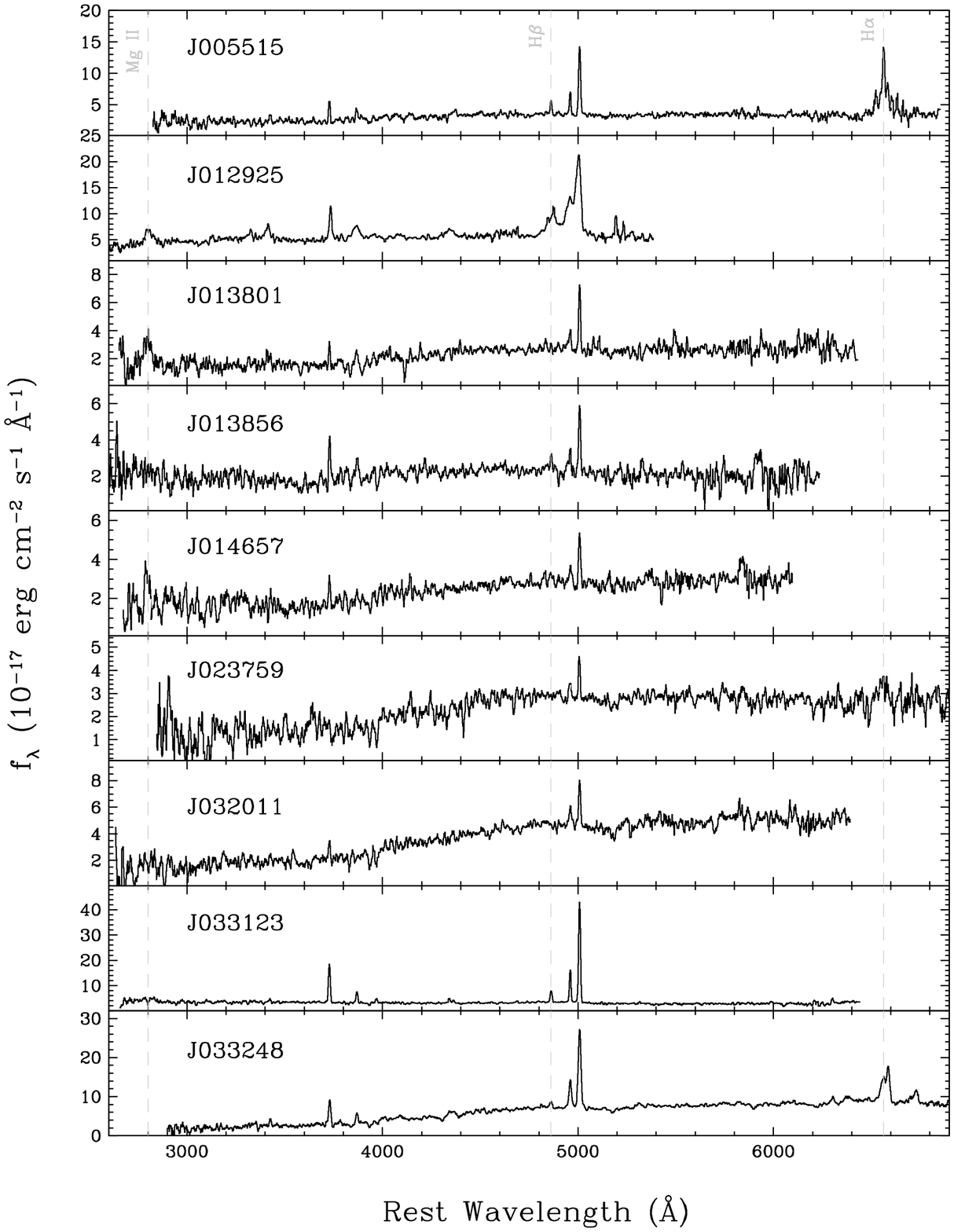}}}
\caption{SDSS spectra of the variable AGNs. Data have been binned to 2
  \AA\ per wavelength bin and smoothed with a 5-pixel boxcar.
\label{spectra1}}
\end{center}
\end{figure*}
  
\addtocounter{figure}{-1}

\begin{figure*}[t!]
\begin{center}
\scalebox{0.6}{\includegraphics{{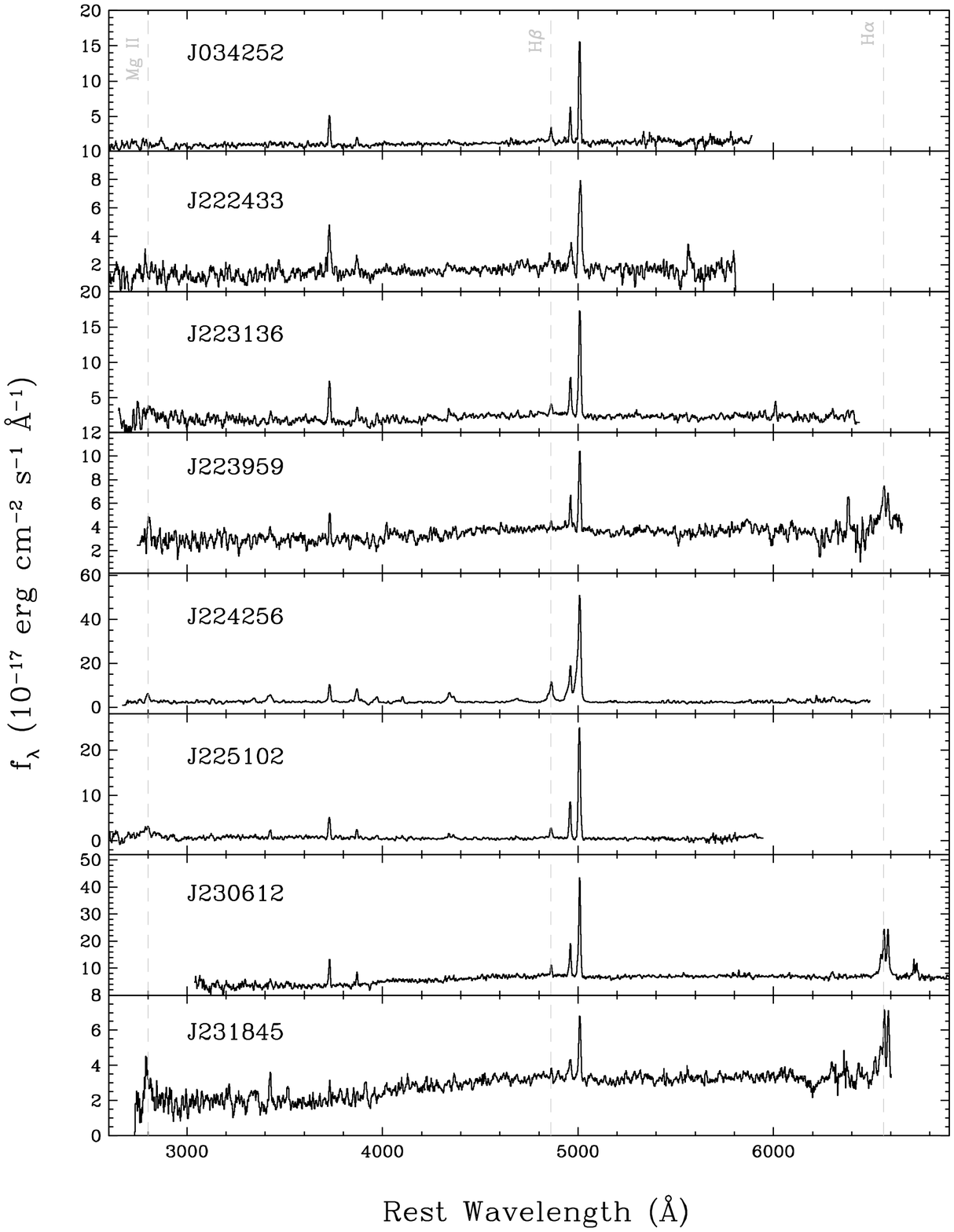}}}
\caption{(Continued)
\label{spectra2}}
\end{center}
\end{figure*}

Light curves of objects within and somewhat beyond the $1\sigma$
scatter band appear non-variable within the measurement uncertainties,
as shown for one sample object in Figure \ref{fig:examplelightcurve}.
The problem of identification of variable sources then becomes a
question of selecting some appropriate threshold above the nonvariable
locus.  From examination of light curves from sources from several
fields, we choose a threshold for selection of variable sources
corresponding to a deviation of $5\sigma$ above the non-variable
locus.  The $5\sigma$ threshold is illustrated by the yellow curve in
Figure~\ref{fig:example}.  We apply this criterion to the fields
containing all 173 Type 2 AGNs.  In selecting this threshold, we have
chosen to prioritize purity over completeness.  In other words, visual
examination of all light curves that exceed the $5\sigma$ threshold
confirms the presence of systematic flux variations that show some
coherence over multiple observation dates, while the same is not true
systematically for lower selection thresholds.  There are likely to be
some genuinely variable objects that lie below the $5\sigma$ cut due
to a weaker level of variability, or variability affecting only a
small portion of the light curve, so our sample of variable AGNs
represents a lower limit to the total number of variable objects in
the Z03 and R08 parent samples. The number of selected variable AGNs
is not very sensitive to small changes in the detection threshold. If
we change the threshold by $\pm0.3\sigma$ for example, the change in
the number of selected variable sources is $\pm1$.

Each field is analyzed separately, as the different photometric
characteristics of the dataset for each field result in different
locations for the non-variable locus and the $5\sigma$ variability
threshold.  In particular, the location of the non-variable locus and
the scatter of objects around it depend primarily on the mean seeing
and the scatter in seeing for all of the observations of a given
field.  The overall picture of the variability selection is shown in
Figure~\ref{fig:var_src}.  In this figure, the non-variable locus and
$5\sigma$ threshold for each field are illustrated as gray and yellow
curves, respectively.  It is apparent that there is a wide range in
the variability thresholds for different fields, due to the variations
in image quality.  Red squares show the Z03 and R08 AGNs lying above
the $5\sigma$ threshold for each field, and blue circles illustrate
the AGNs falling below the $5\sigma$ cut, while the small black points
illustrate \emph{all} detected sources that satisfy the variability
selection in each field.

The majority of the Type 2 AGNs fall very close to the non-variable
loci in the diagram, consistent with the general expectation that Type
2 AGNs should not be strongly variable in the optical.  However, there
are 17 objects from the Z03 and R08 catalogs which lie above the
$5\sigma$ thresholds for their respective fields. Their SDSS
designations, redshifts, median magnitudes, rms variability amplitudes
(\sigmasixtyeight), and [\ion{O}{3}] luminosities are given in
Table~\ref{tab:var_agns} and we discuss their properties below.  We
find 13 variable AGNs from Z03 and 6 variables from R08, with two
objects in common between the two samples.  The light curves of the
variable sources are displayed in Figure \ref{fig:lightcurves}.

Our variability selection method based on rms brightness fluctuations
allows detection of objects showing significant flux changes over and
above the level expected from photometric measurement errors alone,
but it gives no information about the timescale over which variations
occur.  The measured value of \sigmasixtyeight\ for an AGN depends on
the underlying variability behavior, the cadence of observations, and
the photometric quality of the images which sets the level of random
measurement errors.  Thus, \sigmasixtyeight\ combines variability
information on all observed timescales and uses the full available
information in the light curve to test for variability.  By
determining the variability threshold separately for each Stripe 82
field, each AGN is compared with other objects that were observed with
the same cadence as well as identical observing conditions at each
epoch; this provides a consistent framework for detection of variable
sources despite the wide range in variability thresholds in different
fields (Figure \ref{fig:var_src}).

Figure \ref{spectra1} illustrates the SDSS spectra of the selected
variable sources.  Close examination of the spectra shows that many of
them contain definite or possible broad emission lines in their
spectra, primarily either \hal\ or \ion{Mg}{2}.  In some cases the
identification of broad emission lines appears plausible but
ambiguous, because \hal\ and \ion{Mg}{2} appear at the noisy red or
blue ends of the spectra. Among these 17 AGNs, one object is easily
identified as a misclassified Type 1 quasar: J012925 (from the R08
sample) has broad \hbeta, \hgamma, \hdelta, and \ion{Mg}{2}. It may
have been misidentified as a Type 2 AGN in the R08 sample because it
has a peculiar spectrum with unusually broad [\ion{O}{3}] emission, in
which the 4959 \AA\ and 5007 \AA\ lines are strongly blended. The
individual spectra are described in Section \ref{individualobjects}.
In most cases, a higher S/N spectrum would be needed to carry out a
definitive test for broad-line emission, but even with only the SDSS
spectra it appears plausible that some of these variable objects are
actually Type 1 AGNs.

\section{Spectropolarimetry}
\label{sec:spol}

In order to distinguish among possible explanations for the observed
variability, we conducted new observations using the LRISp dual-beam
spectropolarimeter \citep{oke1995,goodrichlris1995} at the Keck I
telescope.  With these observations, our aims were to test for the
presence of broad emission lines in the total flux spectrum using
spectra of broader wavelength coverage and higher S/N than the SDSS
spectra, and to search for strongly polarized continua and/or
polarized broad emission lines that could provide evidence for the
existence of highly obscured nuclei seen only in scattered light. One
possible outcome would be a measurement of low polarization, together
with the detection of broad lines in total flux.  This would imply
that a target is actually a Type 1 AGN that was misclassified as a
Type 2 object, or possibly that it underwent a state change due to a
substantial change in line-of-sight obscuration since the SDSS
spectrum was taken.  Alternatively, a detection of highly polarized
continuum and broad-line emission would indicate that the detected
variability must be coming from the scattered light component,
implying a very compact size (smaller than several light-years) for
the scattering region.  If an object has low polarization and does not
show broad lines in either total flux or polarized flux, it would
remain a candidate for being a naked Type 2 quasar lacking a BLR,
although additional evidence would be required in order to confirm
this classification, such as measurement of the X-ray obscuring column
density.

\begin{figure*}[t!]
\begin{center}
\scalebox{0.6}{\includegraphics{{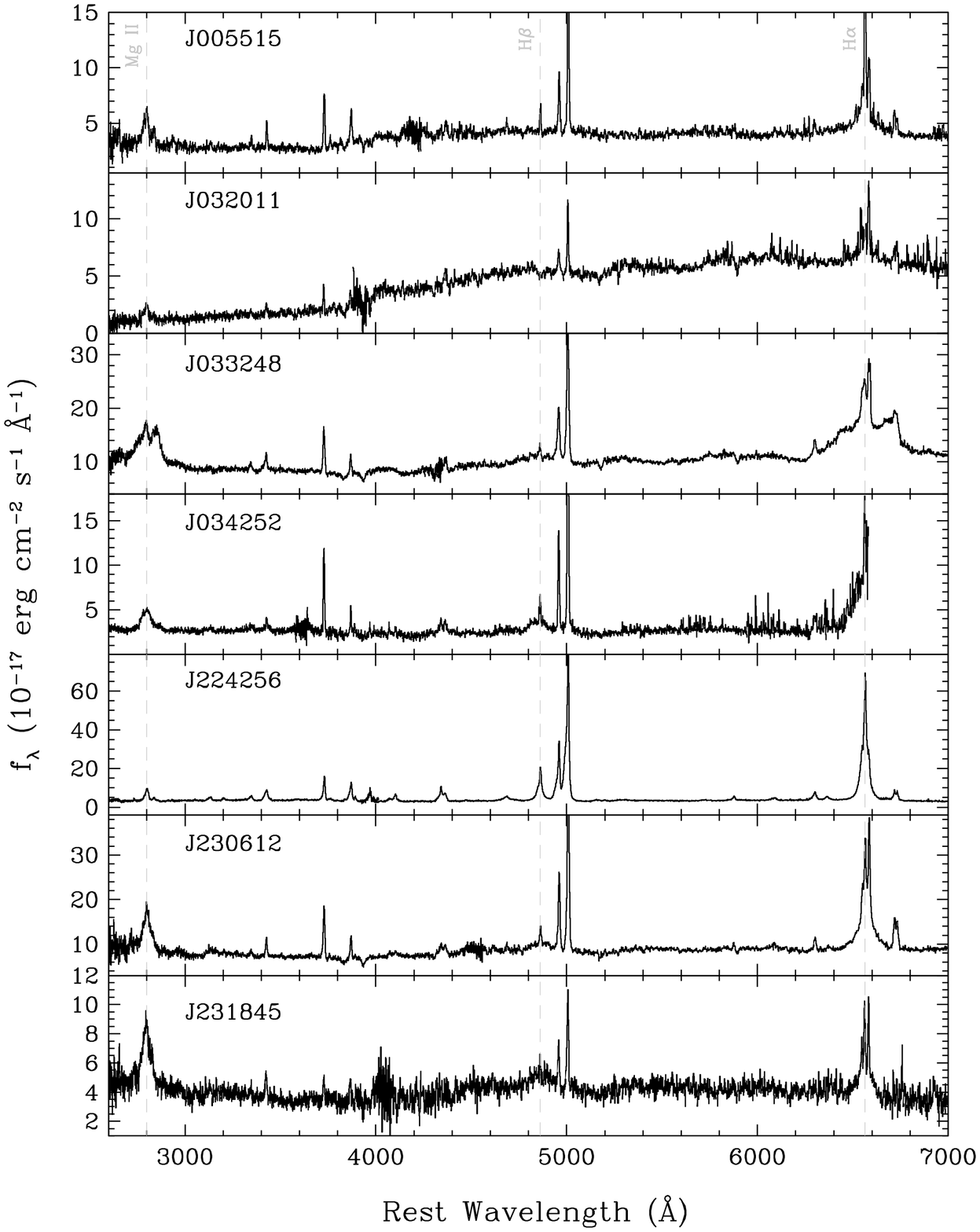}}}
\caption{LRISp total-flux spectra of the seven AGNs observed at Keck.
  The noisy spectral region seen near 4000--4500 \AA\ in some spectra
  corresponds to the dichroic cutoff at 5600 \AA\ in the observed
  frame.
\label{lrisspectra}}
\end{center}
\end{figure*}

The LRISp observations were carried out during the night of 2012
September 20 UT in clear conditions, with seeing around 1\arcsec.  We
used a 1\arcsec\ slit width, and a D560 dichroic to separate the blue
and red beams.  The blue side of the spectrograph used a 400 grooves
mm\per\ grism and the red side used a 400 grooves mm\per\ grating,
giving dispersion of 1.09 \AA/pixel and 1.16 \AA/pixel on the blue and
red sides, respectively.  The total wavelength coverage for the
combined blue and red sides was 3100-10300 \AA, with a small overlap
at 5600--5700 \AA.

During the course of the night, we observed seven AGNs selected from
the list described above. Each was observed in a standard sequence of
four settings of the rotatable half-wave plate (0\arcdeg, 45\arcdeg,
22\fdg5, and 67\fdg5), with exposure times of 680 and 600 seconds for
the blue and red exposures, respectively.  Shorter exposure times were
used for the red side to compensate for the longer readout time of the
LRIS red CCDs.  Each AGN was observed for one full observation
sequence, for a total of 2720/2400 s (blue/red sides), except that for
J224256 we obtained two exposure sequences.  Additionally, we observed
the null (unpolarized) star HD 154892, the polarized standards HD
155528 and HD 283812, and the flux standards BD +28\arcdeg4211 and
G191B2B.  Data were reduced and calibrated following methods described
by \citet{miller1988} and \citet{barth1999}.  Over the wavelength
range 4500--5400 \AA, the null standard star had
$p=(0.038\pm0.003)\%$, indicating a negligible level of instrumental
polarization.

For the polarization measurements described below, we use only the
blue-side data. The red-side data are more impacted by residuals from
strong cosmic-ray hits in the LRIS red detectors, which are 300
\micron-thick LBNL CCDs \citep{rockosi2010}.  The high rate of
cosmic-ray events limited the feasible exposure times to $\sim10$
minutes, which severely constrained the S/N achievable in a single
exposure. Imperfect cleaning of the very large multi-pixel cosmic-ray
hits on the red CCD left numerous small residuals that caused spurious
features in the Stokes parameter spectra.  These residuals are more
severe for spectropolarimetry than for standard spectroscopy because
the polarization calculation relies on measurement of the small flux
differences between the orthogonally polarized beams.  Additionally,
during much of the night the spatial focus in the red camera was
relatively poor and in some of the red-side exposures the spatial
profile of the target was slightly bifurcated; this focus problem did
not affect the blue camera.  As a result, we were able to extract
reasonable total-flux spectra for the red side, but the red
polarization measurements were not as reliable as the blue side data.

Figure \ref{lrisspectra} illustrates the LRISp total-flux spectra of
these seven objects. The primary result that emerges from the LRISp
spectra is that all seven of these objects have broad \hal\ and/or
broad \ion{Mg}{2} emission lines in total flux.  These broad lines are
fairly obvious in five of the seven objects: J005515, J033248,
J034252, J230612, and J231845.  In the case of J032011, the S/N in
the LRISp spectrum is fairly low, but close examination clearly shows
the \ion{Mg}{2} and \hal\ lines to be much broader than [\ion{O}{3}]
$\lambda5007$.  The remaining AGN, J224256, has a spectrum which might
appear superficially to be consistent with a Type 2 classification,
but we find that it has broad \ion{Mg}{2} emission as well as broad
wings on \hal\ and \hbeta\ that are not present on the forbidden
[\ion{O}{2}] and [\ion{O}{3}] lines. Section \ref{individualobjects}
contains a more detailed discussion of the classification of this
object and others.  Given that our sample selection began with
catalogs of Type 2 AGNs, it is a rather striking result that
\emph{all} of the variable objects for which we have obtained new
spectra show broad emission lines in their total flux spectra.

In no case do we see any evidence for features in the Stokes parameter
spectra at the wavelengths of permitted emission lines, so we do not
identify any examples of hidden broad line regions in this sample.  We
measured the continuum polarization of each object over a broad
wavelength range on the blue side, selected for each object to exclude
strong emission lines.  Table \ref{poltable} lists the continuum
polarization measurements, as percent linear polarization and
polarization angle $\theta$.   

Four objects have continuum polarizations that are essentially
consistent with zero, and two others have weak polarizations of about
1\%.  The only object in our sample with a polarization greater than
2\% is J224256, with $p = (2.39\pm0.37)\%$ over 4920--5320 \AA.  We
did not observe interstellar probe stars to measure the Galactic
interstellar polarization \citep[e.g.,][]{tran1995}, so we cannot
directly determine the Galactic contribution to the measured
polarization.  However, we can estimate the maximum expected value of
the Galactic interstellar polarization from the reddening along each
line of sight, using the extinction map of \citet{schlafly2011} and
assuming $R_V=3.1$.  According to \citet{serkowski1975}, the upper
bound to the interstellar polarization is
$p_\mathrm{max}(\%)=9E(B-V)$, while a typical value for interstellar
polarization is $\sim3E(B-V)$. With a Galactic reddening of
$E(B-V)=0.07$, the expected foreground interstellar polarization for
J224256 is $\sim0.2\%$ with a likely maximum value of 0.6\%.  Thus,
the detected $\sim2\%$ continuum polarization cannot be entirely
ascribed to Galactic interstellar polarization, and some polarization
from the object itself is required, either from the AGN itself, from
absorption or scattering within its close environment, or from
interstellar absorption within the host galaxy.  At the wavelengths of
broad emission lines, the polarization is indistinguishable from the
continuum polarization: for \ion{Mg}{2} we find $p=(2.5\pm0.8)\%$ over
3940--4000 \AA\ (in observed wavelength), and for \hbeta\ we find
$p=(2.2\pm0.3)\%$ (integrated over 6830--6960 \AA).\footnote{At
  wavelengths $\gtrsim9000$ \AA\ the polarization calibration becomes
  unreliable and we are unable to measure a robust polarization for
  the \hal\ line.}  Thus, there are no polarization features that
would suggest the presence of a hidden broad-line region in this
galaxy.

These results stand in marked contrast to those of
\citet{zakamska2005}, who found much higher blue continuum
polarizations of 3--17\% in 9 out of 12 objects observed from the Z03
sample.  For our sample, the polarization characteristics are much
more consistent with the properties of Seyfert 1 galaxies, in which
optical polarizations of $\lesssim1\%$ are typical and $p$ rarely
exceeds 2\% \citep{smith2002}.  This suggests that the variability
selection is picking out a subset of the parent sample having
properties that are significantly different from the majority of Type
2 quasars that have been previously chosen for follow-up observations.

\begin{deluxetable*}{lcccc}
\tablecaption{Continuum Polarization}
\tablehead{
  \colhead{Object} &
  \colhead{$E(B-V)$ (mag)} &
  \colhead{$\lambda_\mathrm{obs}$ (\AA)} &
  \colhead{$p$ (\%)} &
  \colhead{$\theta$ (deg)}
}
\startdata
J005515 & 0.03 & 4000--4900 & $0.59\pm0.48$ & $155\pm15$ \\
J032011 & 0.09 & 4200--5300 & $1.03\pm1.40$ & \nodata \\
J033248 & 0.10 & 4000--4800 & $1.38\pm0.20$ & $175\pm4$ \\
J034252 & 0.11 & 4600--5200 & $1.16\pm0.51$ & $133\pm11$ \\
J224256 & 0.07 & 4920--5230 & $2.39\pm0.37$ & $115\pm4$ \\
J230612 & 0.05 & 4000--5300 & $0.07\pm0.18$ & \nodata \\
J231845 & 0.04 & 4000--5300 & $0.05\pm0.80$ & \nodata \\
\enddata
\tablecomments{Wavelength ranges for the polarization measurement are
  listed in the observed frame.  
Polarization angles are not reported for objects having
continuum polarization consistent with zero.  Galactic foreground
reddening $E(B-V)$ values are based on \citet{schlafly2011} and taken from
NED, assuming $R_V=3.1$. The maximum expected Galactic interstellar
polarization is $p_\mathrm{max}(\%) \approx 9E(B-V)$
\citep{serkowski1975}.  }
\label{poltable}
\end{deluxetable*}

\section{Notes on Individual Objects}
\label{individualobjects}

Below, we describe features of the SDSS and Keck spectra and discuss
the classification of each variable object.

\emph{SDSS J005515.81$-$004648.5:} The SDSS data show the likely
presence of a broad \hal\ line at the red end of the spectrum,
although it is very noisy. The LRISp spectrum confirms the presence of
broad \hal\ wings as well as a broad \ion{Mg}{2} line.  No broad wings
are apparent on the \hbeta\ line, so this object would be best
classified as a Type 1.9 AGN.

\emph{SDSS J012925.82$-$005900.2:} This object has a definite Type 1
AGN spectrum, although it is unusual in that the [\ion{O}{3}]
$\lambda\lambda4959,5007$ lines are fairly broad and blended together.
Broad lines in the spectrum include \hbeta, H$\gamma$, H$\delta$, and
\ion{Mg}{2}. This object is included in the SDSS DR5 quasar catalog
\citep{schneider2007}.\footnote{The SDSS DR8 archive server
  incorrectly assigns a redshift of $z=6.03$ to this object, due to
  misidentification of [\ion{O}{3}] as Ly~$\alpha$.}

\emph{SDSS J013801.57$-$004946.5:} Broad \ion{Mg}{2} $\lambda2800$
is present at the noisy blue end of the SDSS spectrum.

\emph{SDSS J013856.14+003437.4:} The source has fairly weak
emission lines, and the S/N of the SDSS spectrum is too low to
clearly distinguish whether a broad \hbeta\ component is present. It
lies just barely above the $5\sigma$ variability threshold for
inclusion in our sample.

\emph{SDSS J014657.23+005537.2:} \ion{Mg}{2} is visible in the
SDSS spectrum and appears broader than [\ion{O}{2}] or [\ion{O}{3}],
but the S/N is fairly low.

\emph{SDSS J023759.75+001723.5:} This source has a
starlight-dominated spectrum with weak emission lines and possible
weak broad bump at the wavelength of \hal.

\emph{SDSS J032011.94+002702.2:} This is another starlight-dominated
source with weak emission lines. Broad \ion{Mg}{2} is likely present
at the blue end of the SDSS spectrum, and confirmed in the LRISp
data. Broad \hal\ wings are also visible in the LRISp spectrum,
although the red-side data are noisy and impacted by night-sky
emission-line residuals.

\emph{SDSS J033123.14$-$005930.7:} This source has strong narrow
emission lines and does not appear to have broad wings on the
\hbeta\ emission line.  However, a broad, very low-amplitude
\ion{Mg}{2} $\lambda2800$ line is possibly present.

\begin{figure}
  \includegraphics[width=\columnwidth]{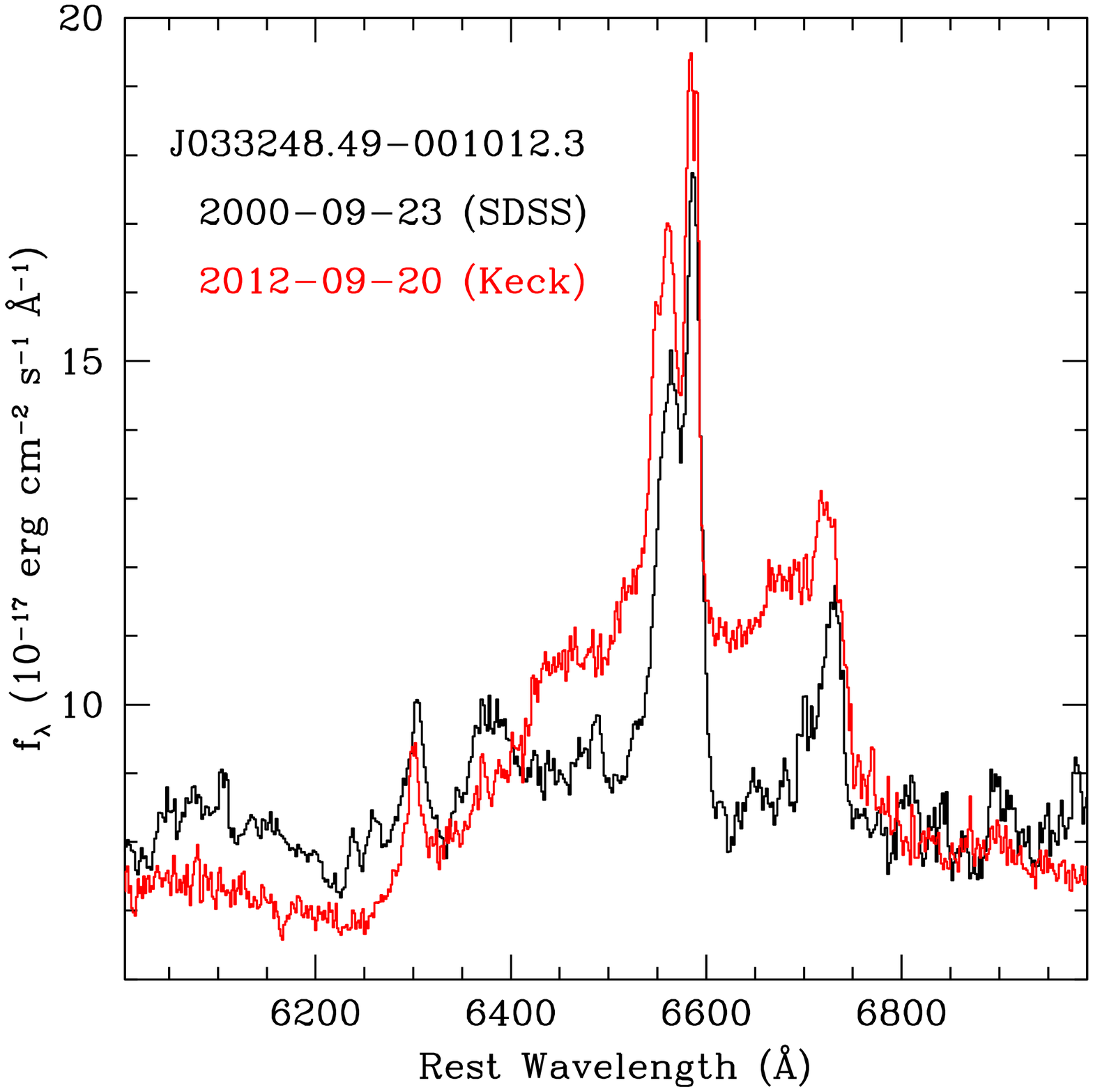}
  \caption{Comparison between the SDSS spectrum (smoothed by a 5-pixel
    boxcar) and the LRISp total flux spectrum (scaled by a factor of
    0.67) for SDSS J033248.49$-$001012.3, illustrating the change in
    the broad \hal\ profile over a 12-year span.
    \label{fig:j033248}
  }
\end{figure}

\emph{SDSS J033248.49$-$001012.3:} The SDSS spectrum initially appears
consistent with a Type 2 AGN classification, with a
starlight-dominated continuum. However, the LRISp spectrum reveals a
strong and extremely broad \hal\ emission line as well as broad
\ion{Mg}{2}, giving an appearance similar to typical double-peaked
\hal\ emitters \citep[e.g.,][]{strateva2003}. Comparison between the
SDSS and Keck spectra, taken 12 years apart, shows that the very broad
\hal\ feature was probably present but much weaker in the earlier data
(Figure \ref{fig:j033248}). The SDSS spectrum has a prominent bump at
rest wavelength $\sim6400$ \AA\ and an extended ``shelf'' blueward of
the narrow \hal\ emission line, and the AGN was probably classified as
a Type 2 quasar as a result of the relatively low amplitude of these
broad features, as well as the lack of an obvious ``normal''
broad-line component.  This object is also included in the SDSS quasar
catalog \citep{schneider2007} and has been considered as a Type 1 AGN
in other studies \citep{shen2008}.

Of the seven objects having spectropolarimetry data, J033248 is the
only one detected in the FIRST radio survey \citep{becker1995}, where
it is found to be a 15.7 mJy source.  The deconvolved major and minor
axis sizes are 1.77 and 0.91 arcsec, respectively, with the major axis
oriented at PA=179\fdg6, nearly parallel to the optical polarization
angle.  \citet{smith2004} interpret the relative orientation of the
radio and optical polarization position angles in Type 1 AGNs as a
consequence of the inclination angle of the AGN.  In this picture
\citep[see also][]{marin2012}, objects in which the optical
polarization angle is parallel to the radio axis are viewed at
relatively face-on inclinations ($<45\deg$) and dominated by
equatorial rather than polar scattering.  The properties of this
object appear consistent with this interpretation, although we caution
that the observed polarization of $1.38\%$ is only slightly larger
than the maximum allowed Galactic interstellar polarization of
$\sim1.0\%$, which leaves open the possibility that the relative
alignment of the radio and optical polarization axes could be
substantially affected by the Galactic contribution.

\emph{SDSS J034252.47+005252.4:} The SDSS spectrum appears
consistent with a Type 2 classification, but the LRISp spectrum
reveals broad \ion{Mg}{2}, broad \hbeta\ wings, and the blue half of a
broad \hal\ line, cut off at the end of the red spectrum.  This AGN
appears in the SDSS DR5 quasar catalog \citep{schneider2007}, and is
included in both the Z03 and R08 samples.

\emph{SDSS J222433.31$-$003634.0:} The SDSS spectrum is fairly
noisy but appears consistent with a Type 2 classification.

\emph{SDSS J223136.27$-$011045.0:} The SDSS spectrum shows a noisy
broad bump at \ion{Mg}{2}.

\emph{SDSS J223959.04+005138.3:} A broad \hal\ line is likely
present but is in the very noisy region at the red end of the SDSS
spectrum.

\begin{figure}
  \includegraphics[width=\columnwidth]{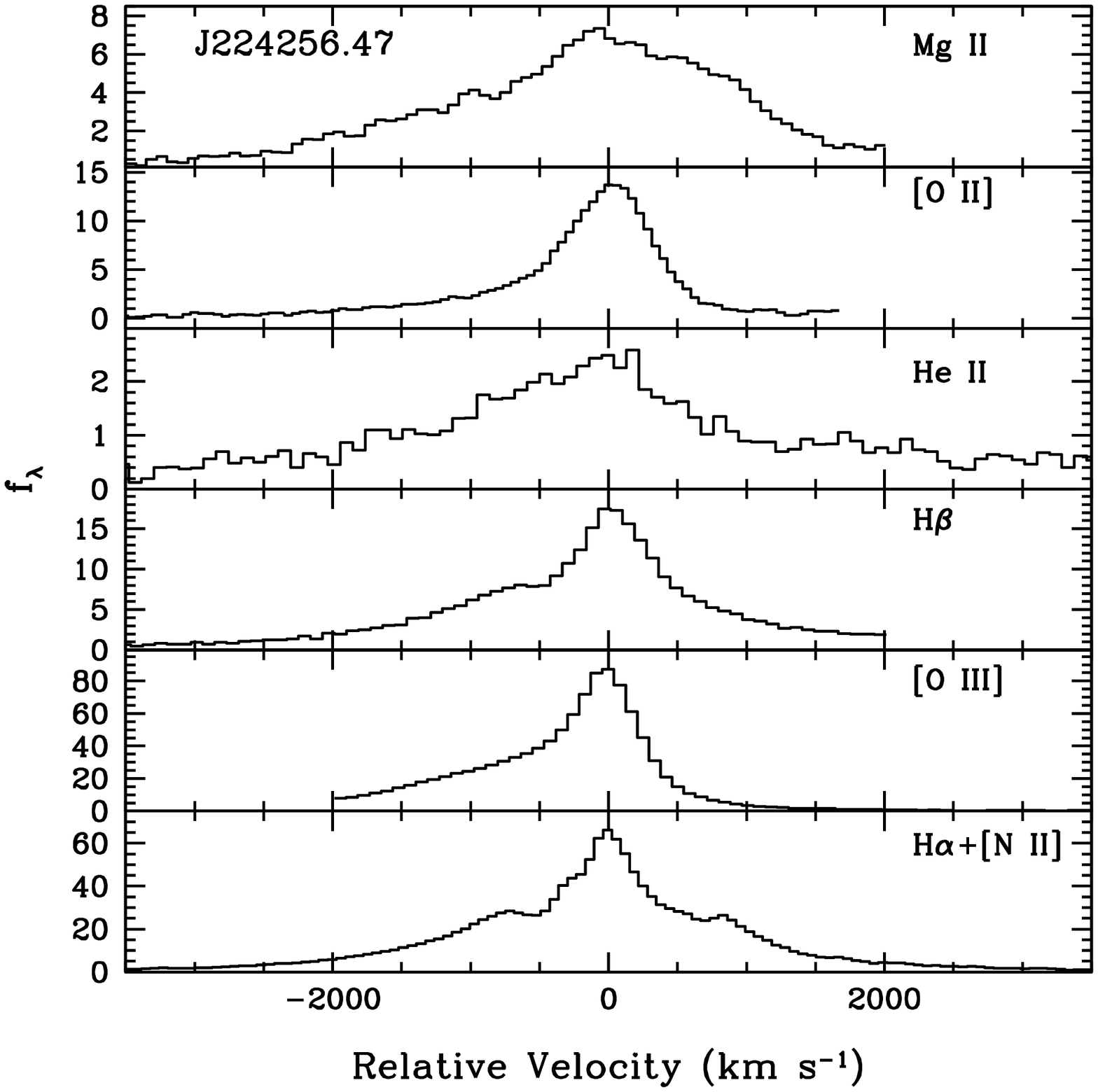}
  \caption{Comparison of line profiles for SDSS J224256.47+005155.2 in
    the Keck LRISp spectrum, illustrating broad permitted emission
    features as well as extended blue wings on the narrow emission
    lines.
    \label{fig:j224256profiles}
  }
\end{figure}

\emph{SDSS J224256.47+005155.2:} The SDSS spectrum shows strong
narrow emission lines and a featureless continuum. Broad, blueshifted
wings are present on the [\ion{O}{3}] lines, while \hbeta\ appears to
have an extended red wing not present on [\ion{O}{3}].  A weak,
low-amplitude broad \ion{He}{2} $\lambda4686$ emission line is also
visible. Overall, the SDSS spectrum appears to be plausibly consistent
with a Type 1 classification, albeit with relatively narrow and weak
broad lines, and this source is included in the SDSS DR5 quasar
catalog \citep{schneider2007}. Figure \ref{fig:j224256profiles}
illustrates the profiles of several emission lines in the Keck LRISp
spectrum.  Although the blueshifted wings of [\ion{O}{2}] and
[\ion{O}{3}] extends out to velocities beyond 2000 \kms, the Balmer
lines have red wings extending to similarly high velocities while the
forbidden oxygen lines do not. The contrast between the widths of
\ion{Mg}{2} (FWHM $\approx 2400$ \kms) and [\ion{O}{2}] (FWHM $\approx
880$ \kms) gives the clearest indication that this is a Type 1 AGN.

Interestingly, SDSS J224256 has a spectrum that is overall extremely
similar to that of IRAS 20181$-$2244, an object at $z=0.185$ that was
originally classified as a Type 2 quasar by \citet{elizalde1994}. It
was later demonstrated to be a Type 2 impostor by \citet{halpern1998b}
who detected broad wings on the Balmer lines and reclassified it as a
narrow-line Seyfert 1 (NLS1) galaxy.\footnote{\citet{halpern1998b}
  refer to IRAS 20181$-$2244 as a ``I~Zw~1''-type object because of
  the narrowness of its permitted broad lines, although it does not
  have the extremely strong \ion{Fe}{2} emission that is a
  characteristic of I~Zw~1 itself. Such objects are most often
  referred to as NLS1s, but they avoided that term in order to prevent
  confusion with narrow-line (i.e., obscured, Type 2) AGNs. Regardless
  of the terminology, SDSS J224256.47 and IRAS 20181$-$2244 have very
  similar spectra with definite broad Balmer-line components and other
  features that are consistent with a Type 1 classification.}

\emph{SDSS J225102.40$-$000459.8:} There is a low-amplitude,
broad \ion{Mg}{2} line in the SDSS spectrum.

\emph{SDSS J230612.90$-$005912.6:} The spectrum contains strong
narrow emission lines on a starlight-dominated continuum, giving the
appearance of a Seyfert 2 galaxy. However, weak broad wings appear to
be present on the \hal\ line, and this object was included in the
\citet{greene2007} catalog of Type 1 AGNs with black hole mass
estimates based on the broad \hal\ width. Broad \ion{Mg}{2} and
\hal\ wings are obvious in the Keck spectrum, and a low-amplitude
broad \hbeta\ feature is also visible.

\emph{SDSS J231845.12$-$002951.4:} Broad \ion{Mg}{2} and
\hal\ are likely present at the blue and red ends of the SDSS
spectrum, and are confirmed in the Keck spectrum.

\section{Discussion}
\label{sec:discussion}

As described in the Introduction, the simplest version of unified
models for AGNs leads to a general expectation that optical
variability should not be seen in Type 2 AGNs on timescales of a few
years or less. While it is true that 90\% of our SDSS sample is
consistent with this prediction, the detection of variability in 10\%
of the Type 2 AGN population is surprising.  We can consider several
possible explanations for this finding.

\emph{``Naked'' or ``true'' Type 2 AGNs:} One possibility is that
optical variability has been detected in unobscured AGNs that do not
possess a BLR, similar to the so-called naked or true Type 2 AGNs
identified by other means.  X-ray spectroscopy and other observations
have revealed a small number of candidates for unobscured Type 2
Seyfert galaxies showing no evidence for the presence of a BLR
\citep[e.g.,][]{pappa2001, boller2003, wolter2005, panessa2009,
  shi2010, tran2011, bianchi2012, gallo2013, miniutti2013}.  The
existence of such objects at very low luminosities can be understood
in the context of disk-wind models that propose a lower bound in
luminosity or in Eddington ratio ($L/L_\mathrm{Edd}$) below which the
BLR would be unable to form \citep{nicastro2000, elitzur2009,
  trump2011}.  Alternatively, \citet{laor2003} proposed a luminosity
threshold below which the BLR would have such small radius and such
high Keplerian velocities that it could not remain stable.  A more
speculative question is whether unobscured Type 2 AGNs may exist at
quasar-like luminosities (as proposed by Hawkins 2004) or at Eddington
ratios well above the threshold at which the BLR is capable of
forming.  The existence of luminous, unobscured Type 2 AGNs above
$L/L_\mathrm{Edd}\gtrsim10^{-2}$ would require a different physical
explanation than those proposed in the above scenarios
\citep{gallo2013, miniutti2013}.  \citet{wang2012} recently described
a time-dependent model in which the BLR is generated episodically from
a star-forming accretion disk, and they predict that $\sim1\%$ of
luminous AGNs would be observed during transient phases of weak or
absent BLR emission, appearing as true Type 2 objects.  More
observational and theoretical effort is certainly needed to assess the
possible existence of unobscured Type 2 quasars.  However, in our SDSS
sample, unobscured Type 2 AGNs are strongly disfavored as a primary
explanation for optical variability, because the Keck observations
revealed broad emission lines in each of the seven objects that were
observed.  High S/N spectroscopy of the remainder of our sample would
be useful in order to test whether any other objects might still be
candidate true Type 2 AGNs.

\emph{Weak-lined quasars:} Surveys have detected a small population of
unobscured AGNs whose broad emission lines are extremely weak
\citep[e.g.,][]{mcdowell1995, fan1999, diamond2009}.  The relationship
between these weak-lined quasars and true Type 2 AGNs is not entirely
clear, partly because of the differences in selection methods.
Weak-lined quasars have mainly been found at high redshift by
identifying objects with a Lyman break but very little Ly~$\alpha$ or
\ion{C}{4} emission, while true Type 2 AGN candidates are primarily
low-redshift objects identified via their optical narrow-line
emission.  Could low-redshift counterparts of weak-lined quasars
appear as Type 2 quasar impostors? \citet{shemmer2010} presented
infrared spectra of two high-$z$ weak-lined quasars, finding that they
had extremely weak [\ion{O}{3}] emission as well as very weak broad
\hbeta\ emission.  If weak-lined quasars at lower redshift had
similarly weak [\ion{O}{3}] emission, they probably would not be
included in optically-selected Type 2 quasar samples identified by
their narrow-line emission. Although the data available to make this
comparison are very sparse, the rest-frame optical spectra of
high-redshift weak-lined quasars do not appear closely similar to the
Z03 or R08 AGNs.

\emph{Obscured AGNs visible in scattered light:} Obscured AGNs with
hidden central engines visible in scattered light could exhibit some
degree of variability, although the variability signal seen in
scattered light would most likely be weak and temporally blurred by
scattering over a spatially extended mirror.  In this case, the
timescale for significant variability could be used to constrain the
size of the scattering region.  \citet{zakamska2005} examined
\hst\ images of three Type 2 quasars, finding evidence of
kiloparsec-scale, conical scattering regions. If these sizes are
typical of scattering regions in Type 2 quasars, then it would be
unlikely that variability would be detected in the scattered light on
timescales of a few years. We did not detect high polarization in any
of the seven objects observed at Keck, and the spectropolarimetry
results strongly disfavor the possibility that the detected
variability in these AGNs is coming from a scattered light component.

\emph{Changing appearance due to variable continuum luminosity:}
Intermediate-type Seyferts (types 1.8 and 1.9) represent a ``mixed
bag'' of AGNs having relatively weak broad emission lines due to
either a low ionizing continuum flux, or extinction toward the BLR, or
possibly both.  \citet{trippe2010} find that about half of a sample of
local intermediate-type AGNs are classified as such due to a weak
ionizing continuum. In such objects, significant fading of the
continuum could make the BLR fade below detectability, giving the
appearance of a Type 2 AGN but without significant obscuration.  One
example of this behavior is the highly variable AGN NGC 2992
\citep{trippe2008}. Over the past few decades, this object has changed
its appearance from Seyfert 1.5 to 1.9 to 2 primarily as a result of
strong variations in the continuum luminosity, and not due to variable
obscuration. (In a low state, such an object could be considered as an
example of a temporarily naked Type 2 AGN.)

\emph{State changes due to variable obscuration:} If an AGN was
temporarily in a high-obscuration state when the SDSS spectrum was
observed but the line-of-sight obscuration changed substantially
during some portion of the Stripe 82 survey, the object could appear
as a variable Type 2 AGN based on the available data. However, there
are very few documented examples of AGNs changing from fully-obscured
Type 2 states to unobscured Type 1 states; a possible example was
presented by \citet{aretxaga1999}.  More commonly, spectral
variability in Type 1.8/1.9 AGNs due to variable obscuration can lead
to dramatic changes in emission-line strength over timescales of years
\citep{goodrich1989,goodrich1995}.  \citet{halpern1998a} discuss the
instructive case of 1E~0449.4$-$1923, which had previously been
considered to be a Type 2 quasar candidate. They detected strong broad
emission lines in this object and argued that the object was not a
Type 2 quasar at all, but a partially obscured Type 1.8/1.9 object
that emerged as a more obvious Type 1 AGN as a result of decreasing
extinction toward the central engine.  Such ``variable
intermediate-type AGNs'' can easily be mistaken for Type 2 AGNs if
they are observed when the obscuration is relatively high.

\emph{Misclassification due to low S/N or incomplete spectral
  coverage:} The most trivial possibility would be that some of the
variable Type 2 AGNs are simply Type 1 AGNs that were previously
misclassified as a result of inadequate data quality in the original
SDSS spectra, or some shortcoming in the sample selection procedure.
Among the variable sources identified in this paper, only one object
(SDSS J012925.81$-$005900.2) is an example of an obviously mistaken
classification, but at least a handful of other objects seem to have
probable or possible broad \hal\ or \ion{Mg}{2} features in the SDSS
spectra that are too noisy to distinguish securely. The possibility of
misclassification is exacerbated if an AGN happens to be in a faint
state when observed spectroscopically, either due to variable
obscuration or intrinsic flux variability. However, with sufficiently
high S/N and spectral coverage that includes \hal\ and/or \ion{Mg}{2},
it should be possible to distinguish Type 1.8/1.9 AGNs from fully
obscured Type 2 objects. In the past, some purported Type 2 quasars
were found to be Type 1 objects after better data were obtained
\citep{halpern1998b,akiyama2002,panessa2009,gliozzi2010,shi2010}.  The
example of IRAS 20181$-$2244 \citep{halpern1998b} illustrates how
tricky classification can be (similar to SDSS J224256.47+005155.2
discussed in this paper), because there are subclasses of Type 1 AGNs
even at high luminosity that have very narrow permitted lines and can
easily be mistaken for Type 2 objects.

\emph{Other transient sources:} A supernova in a Type 2 AGN host
galaxy could result in a spurious detection of AGN variability.
However, such an event should in principle be identifiable as a
supernova by its light curve shape, even if no contemporaneous
spectrum was obtained. None of our detected variable sources has a
light curve that appears consistent with a single supernova event.

For the sample discussed in this paper, the new Keck data easily
clarify the situation. Out of seven objects observed at Keck, all
seven are seen to be Type 1 AGNs, and the natural conclusion is that
they were misclassified as Type 2 AGNs primarily because of low S/N
and inadequate spectral coverage in the SDSS data.  It seems likely
that at least in some cases a contributing factor was that some AGNs
may have been in a relatively faint state when the SDSS spectra were
observed. For example, J033248 is seen to have very strongly
double-peaked \hal\ emission in the new Keck spectrum, but this
feature is at most weakly present, if at all, in the earlier SDSS
spectrum.  Among the ten galaxies not having new Keck data, one is
definitely a Type 1 AGN, and we identify possible weak broad emission
lines in seven others based on examination of the SDSS data.  Given
that the Keck data had a 100\% success rate for confirmation of
possible or ambiguous broad lines that were seen in the SDSS data, it
seems reasonable to extrapolate our results to conclude that the vast
majority (and possibly all) of the variable sources are actually Type
1 AGNs.  Based on the apparent weakness of \hbeta\ in the SDSS and
Keck spectra, most of these appear to be variable
``intermediate-type'' 1.8 or 1.9 objects, similar to low-redshift AGNs
previously identified as showing spectral variability
\citep{goodrich1989, goodrich1995}.

A larger fraction of our variable sources comes from the Z03 catalog
than from the R08 catalog; we identify 13 variable Z03 objects in
Stripe 82, but only six variable sources from the larger R08 sample
(with two variable sources appearing in both samples).  These were
selected from a parent subsample of 89 Stripe 82 light curves for the
Z03 sample, and 119 light curves for objects in the R08 sample
(including 35 objects that overlap between the two).  If we assume
that all of the variable sources are misclassified Type 1 AGNs, this
implies a contamination fraction of $\sim15\%$ for the Z03 sample, and
$\sim5\%$ for the R08 sample. The lower contamination rate for R08
follows from their modified selection criteria, which are more
efficient at excluding objects with weak broad lines. It would be
straightforward to obtain new, high S/N spectroscopy of the
remaining variable sources to test for the presence of broad emission
lines, to determine whether \emph{all} of the variable sources are
Type 1 AGNs.

Overall, our results highlight the difficulty in identifying pure
samples of Type 2 quasars by optical selection methods, echoing
previous work by \citet{halpern1998a}, \citet{halpern1998b},
\citet{halpern1999}, and others.  Secure optical classifications of
Type 2 AGNs require high S/N spectroscopy, and broad wavelength
coverage can be critical. In particular, identifying AGNs as Type 2
objects based solely on \hbeta\ as the only BLR diagnostic line is
particularly risky, in that broad \hbeta\ is often undetectably weak
even in cases where broad \hal\ wings or broad \ion{Mg}{2} are quite
obvious.  Since our results indicate that all or nearly all variable
``Type 2'' quasars in Stripe 82 are actually Type 1 objects, future
surveys could use optical variability as a criterion to exclude Type 1
contaminants from Type 2 AGN samples.

\section{Summary and Conclusions}

We have carried out a search for optical $g$-band variability among
previously identified luminous Type 2 AGNs in SDSS Stripe 82, using
image-subtraction photometry.  Our conclusions can be summarized as
follows.

1) No significant variability is found in the vast majority
($\sim90\%$) of the Z03 and R08 AGNs. This result is broadly
consistent with the conclusions of \citet{yip2009} who found no
evidence of spectral variability in Type 2 AGNs, and is consistent
with straightforward expectations for AGN unification models.

2) Among 173 AGNs for which light curves could be constructed, we
found evidence of significant variability (above our $5\sigma$
threshold) in 17 objects. Inspection of the SDSS spectra revealed
possible or definite broad emission lines in several cases, suggesting
that variability selection may be an efficient method for
identifying Type 1 contaminants in Type 2 AGN samples.

3) Keck spectropolarimetry data were obtained for seven objects, none
of which have high polarization, and all of which show broad emission
lines in total flux.

4) Based on the available evidence,we conclude that most and possibly
all of the variable AGNs in Z03 and R08 samples are actually Type 1
objects (mostly of Type 1.8/1.9) that contaminate Type 2 samples.

5) While we cannot rule out the possibility that a very small number
of these variable AGNs might be unobscured Type 2 objects (``naked''
AGNs), we conclude that it is much more likely that the variability
selection is simply picking out misclassified Type 1
objects. Follow-up spectroscopy of the remaining variable AGNs would
be useful to identify or rule out any remaining candidates for naked
AGNs.

6) These results serve as a reminder of the difficulty in identifying
pure samples of optically-selected Type 2 AGNs at high luminosity,
although the small number of variable sources detected from the R08
catalog suggests a fairly low contamination fraction at a $\sim5\%$
level for this sample.

\acknowledgements

Research by A.J.B.\ and D.J.C.\ is supported by NSF grant
AST-1108835. A.V.\ acknowledges support from the research grant RFFI
12-02-01358. We thank Shawn Thorman for his contributions to a
preliminary version of this project. The work at LANL was supported by
the Laboratory Directed Research and Development program.  We thank
the referee for helpful suggestions that improved this paper.

Funding for the SDSS and SDSS-II has been provided by the Alfred P. Sloan
Foundation, the Participating Institutions, the National Science Foundation,
the U.S. Department of Energy, the National Aeronautics and Space
Administration, the Japanese Monbukagakusho, the Max Planck Society, and the
Higher Education Funding Council for England. The SDSS Web Site is
http://www.sdss.org/.

The SDSS is managed by the Astrophysical Research Consortium for the
Participating Institutions. The Participating Institutions are the American
Museum of Natural History, Astrophysical Institute Potsdam, University of
Basel, University of Cambridge, Case Western Reserve University, University
of Chicago, Drexel University, Fermilab, the Institute for Advanced Study,
the Japan Participation Group, Johns Hopkins University, the Joint Institute
for Nuclear Astrophysics, the Kavli Institute for Particle Astrophysics and
Cosmology, the Korean Scientist Group, the Chinese Academy of Sciences
(LAMOST), Los Alamos National Laboratory, the Max-Planck-Institute for
Astronomy (MPIA), the Max-Planck-Institute for Astrophysics (MPA), New
Mexico State University, Ohio State University, University of Pittsburgh,
University of Portsmouth, Princeton University, the United States Naval
Observatory, and the University of Washington.

Some of the data presented herein were obtained at the W.M. Keck
Observatory, which is operated as a scientific partnership among the
California Institute of Technology, the University of California and
the National Aeronautics and Space Administration. The Observatory was
made possible by the generous financial support of the W.M. Keck
Foundation. The authors wish to recognize and acknowledge the very
significant cultural role and reverence that the summit of Mauna Kea
has always had within the indigenous Hawaiian community.  We are most
fortunate to have the opportunity to conduct observations from this
mountain. This research has made use of the NASA/IPAC Extragalactic
Database (NED) which is operated by the Jet Propulsion Laboratory,
California Institute of Technology, under contract with the National
Aeronautics and Space Administration.

\end{document}